# On Stability, Ancillary Services, Operation, and Security of Smart Inverters

Tareq Hossen and Fahmid Sadeque

***Abstract*— This paper presents some recent trends in the research of grid-interactive inverters. Particularly, this paper focuses on stability, ancillary services, operation, and security of single and multiple inverters in the modern power grid. A grid-interactive inverter performs as a controllable interface between the distributed energy sources and the power grid. High-penetration of inverters in a power distribution system can create some technical challenges on the power quality, as well as voltage and frequency control of the system. In particular, a weak grid can lead to voltage oscillation and consequently instability. Moreover, the power grid is moving toward becoming a cyber-physical system in which smart inverters can exchange information for power marketing and economic dispatching. This puts the inverters at the risk of insecure operations. Hence, security enhancement has become another primary concern. Finally, the grid-interactive inverters are operated proportional power-sharing while operating together with many inverters. Recent research on coordinated operations is also discussed in this paper.**

***Index Terms*—Grid-interactive inverters, stability, full order model, ancillary services, multi-inverter operation, weak-grid, virtual inductance, cybersecurity, enhanced operation.**

## I. Introduction

The modern electric power grid is gradually changing with the introduction of power electronics converter-based distributed energy resources (DERs) such as photovoltaic (PV) arrays, wind turbines, battery energy storage, and natural gas-powered microturbines. Solid-state converters serve as the controllable interface between the distributed sources and the power grid when power is generated from these sporadic dc and ac sources. Multiple stages are usually present in these converters, with an inverter serving as the grid-side stage. These are referred to as grid-interactive inverters, and the decentralized generation units are called distributed generators (DGs) [1]-[3]. Grid-interactive inverters enable high integration of renewable energy sources while also allowing for remote and dynamic control [4]-[6]. In North America, several states have integrated renewable energy resources in their power system planning for the future. Los Angles, California aims to achieve 100% of renewable electricity by 2045 along with aggressive electrification targets for buildings and vehicles [7]. Furthermore, California's total solar power generation is nearly 13%, with certain places generating as much as 25% [8]. In [9], it is reported that California has aimed to generate 50% of its energy from renewable sources by 2030.

The increasing penetration of DGs allows for greater flexibility in power networks, where grid-interactive inverters are vital parts of the modern power network [10]-[11]. Fig. 1. depicts a modern grid-interactive inverter with multiple responsibilities. For instance, smart inverters can be configured to provide ancillary services for the power network during abnormal circumstances, improving power quality. This operation procedure for inverters is called grid-supporting mode [1], [12]. The widespread use of DGs has many advantages, but it also poses new issues in terms of power system stability and reliability [13]-[15]. For example, almost zero inertia of inverter-based DGs leads to low-inertia microgrids. A specific case of the low-inertia grid is the weak grid, in which an inverter could become unstable and, therefore, would have to be disconnected from the system [16]. An inverter could also trip due to internal faults such as faults in the semiconductor devices. As such, inverters should be able to detect internal and external abnormalities and should be equipped with control functionalities to operate effectively in those conditions [17], [18]. Furthermore, the distributed control structure of DGs, can allow the power-sharing between DGs, the energy management, and economic dispatch between DGs. This power marketing and economic dispatching cannot be achieved efficiently without a supervisory control structure requiring communication between a utility operator and inverters [18], [19]. Thus, a grid-interactive inverter becomes a cyber-physical system that includes physical parts, e.g. modules, measurement sensors, signal processors, and data packet communications. This new cyber-physical structure of inverters has numerous benefits but can cause new stability and security risks in the power grid [19]-[21]. For example, the communication-enabled inverters could be at risk of being hacked, thus endangering the inverter's security [22], [23].

This paper intends to provide a systematic overview of the grid-interactive inverters. In addition to this introduction section, the grid-interactive inverters modeling is presented in section II. Section II also discusses the stability analysis under different grid conditions. Section III discusses the ancillary services provided by the grid-interactive inverters. The inverter's security features are discussed in section IV. The works on multiple inverter's coordinated operations are reviewed in section V. Lastly, section VI provides the conclusion and prospects.

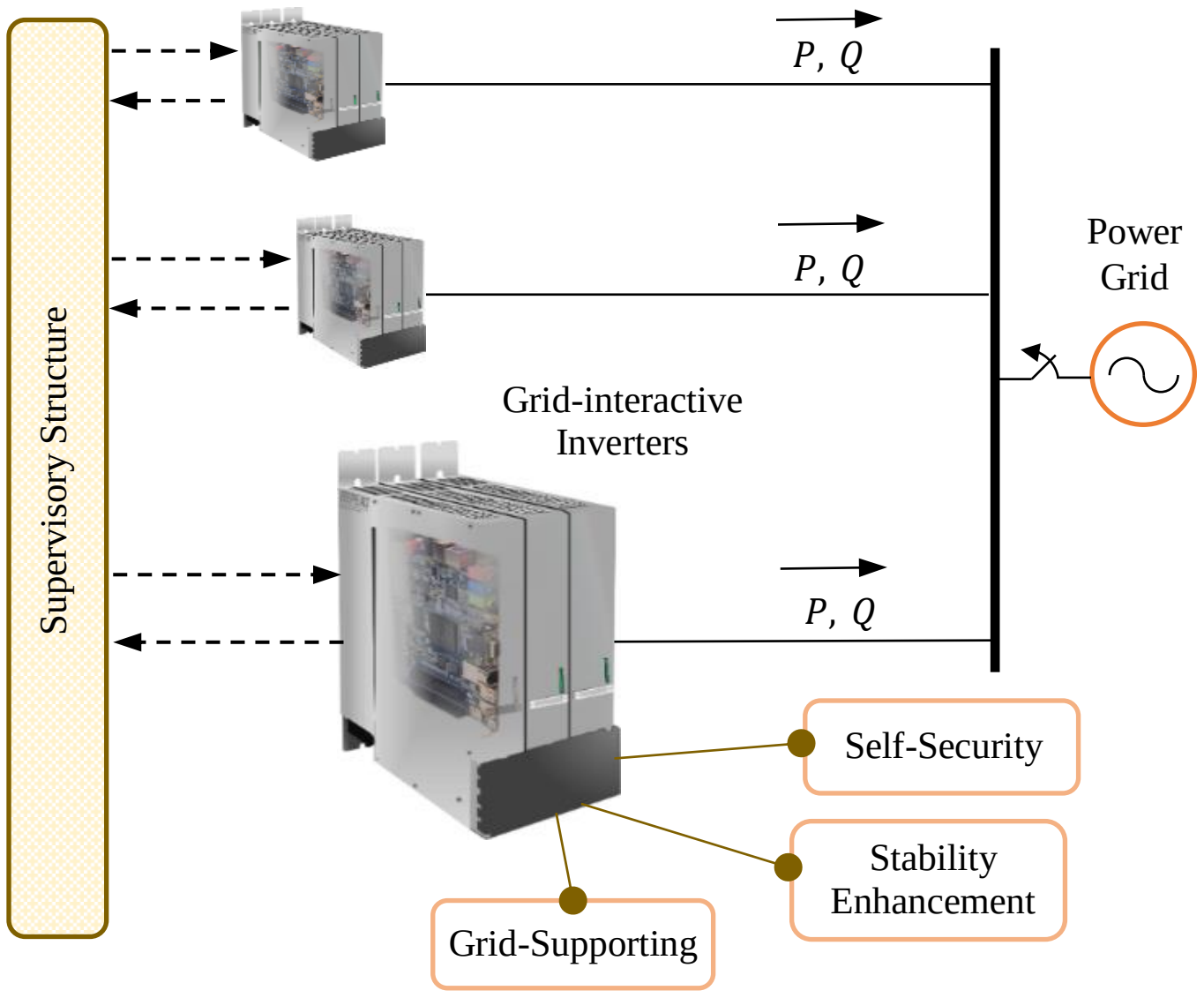

Fig. 1. Diagram of a smart inverter among a network of inverters connected to the power grid.

## II. Modeling and Stability Analysis of Grid-interactive Inverters

Grid-interactive inverters are highly susceptible to grid anomalies. The problem is most noticeable in stand-alone inverter operation and weak grids [24]. A weak grid is formed as power grids having a low short-circuit ratio (SCR), i.e., high impedance and low inertia [25]. Variable impedance in the weak grids can cause unwanted resonance and instability for grid-interactive systems [26]. Additionally, the interface between the controllers and grid fluctuations leads to instability in weak grids [24], [27]. Notably, increasing grid impedances can degrade the performance phase-locked loop (PLL) and negatively affect the grid-interactive system's stability. Furthermore, in [28], it has been shown that the positive feedback gain of anti-islanding methods, a required feature for distributed sources, is limited the stable operation in weak grids. Therefore, it is apparent that weak grids severely degrade grid-tied inverter stability. This section discussed the control and transient stability of grid-interactive inverter systems.

### A. *Modeling of Grid-Interactive Inverter:*

For design and analytical purposes, detailed dynamic models of the grid-interactive inverters are required. However, due to the non-linearity of grid-interactive inverters, i.e., VSI and CSI, the development of dynamic models that can accurately describe their behavior is a complex task. Notice, a VSI is a buck inverter, whereas a CSI can operate as a boost inverter. A control block of grid-interactive VSI is shown in Fig. 2. In CSI, the dc-bus capacitor is replaced by dc-link inductors, $L_{dc}$. The state-space-averaged representation of open-loop grid-connected CSI in the rotating $dq$-frame of reference is derived in [29]-[31]. The state-space-averaged representation of open-loop grid-connected VSI in the rotating $dq$-frame of reference is derived in [24]. The small-signal model of the closed-loop system for VSI and CSI has been formulated in [24], [29]-[33] by replacing the open-loop control inputs with the closed-loop control inputs. These small-signal models have been used to analyze the system stability through eigenvalue trajectories [24], [34]. The stability of the CSI boost inverter with the variation in input signals is studied in [35]-[36]. In [24], the effects of the grid, control scheme, and filter parameters on the stability of PQ-controlled grid-interactive VSI are studied using a twelfth-order state-space model. A full-order inverter dynamic model appropriately represents a DG unit-based system [24]. To perform the stability analysis of a system comprising multiple DG units, a full-order model of the individual unit needs to be determined and then aggregated to represent the actual system. The system model could be complicated because of the integration of several full-order models. Therefore, the detailed inverter model is usually interpreted as a simplified reduced-order model by substituting the inverter by an ideal voltage source (for VSI) or current source (for CSI), avoiding the output filter, and neglecting the grid impedance. One way of eliminating the complexity of a detailed model is to present each inverter by a reduced-order dynamic model resulting from its full-order model. It is important to note that the developed reduced-order model should capture the key feature of the actual system while considerably reducing computational difficulties. In [24], [37], a second reduced-order model of a VSI has been developed to represent actual system characteristics.

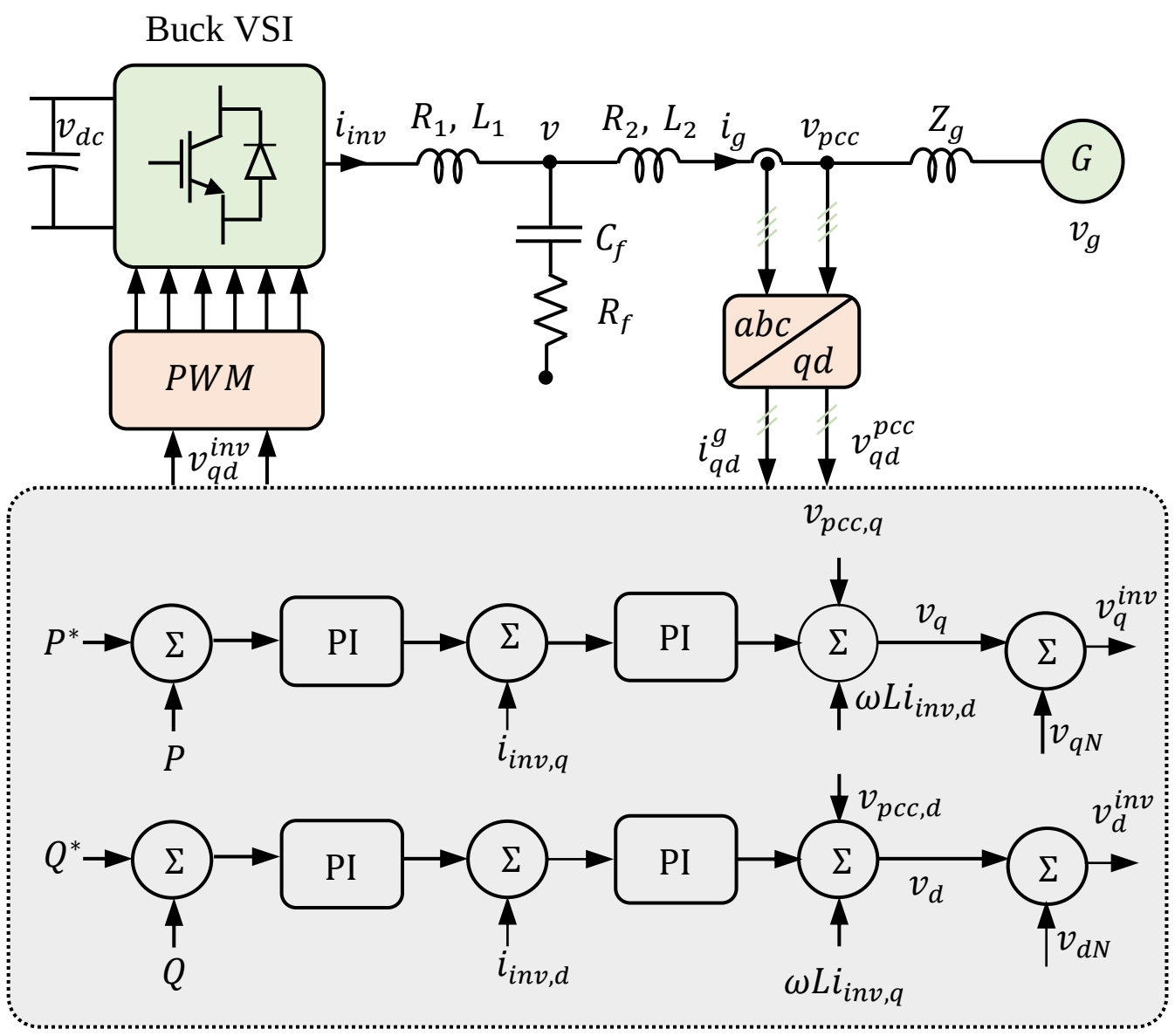

Fig. 2. Control block diagram of a PQ-controlled grid-interactive VSI.

### B. *Stability Enhancement in Weak Grids*

Grid-interactive inverters should be equipped with stability improvement features for operation in weak grid conditions. According to the IEEE Std. 1547.7-2013, a weak grid is characterized by a large impedance between the PCC of the inverter and its primary voltage source, e.g., a substation [24]. Along with the grid impedance, the severity of the weak grid condition is also affected by the size of the DG units compared to the size of the distribution network [36], and also islanded microgrids are particularly vulnerable to weak grids. A voltage fluctuation can be occurred due to the weak grids. When a weak grid condition passes a specific limit, an unstable operation can be observed, resulting in sudden inverter disconnection [36]-[39]. A cascaded failure event can also be observed as a consequence of sudden inverter disconnection [40]-[42]. Therefore, inverter controllers should be equipped with appropriate features to handle the stability issues associated with weak grids. Different stabilizing method for small-signal stability enhancement for grid-interactive inverter in week grids is summarized in Table I.

TABLE I
STABILITY ENHANCEMENT OF INVERTER IN WEEK GRIDS

| **Stabilizing method** | **Proposal** |
|---|---|
| Feedforward techniques | • Virtual impedance [49], [50]<br>• Delay compensation [46]<br>• Feed forward control from PLL [51] |
| Adaptive Techniques | • Adaptive virtual impedance [50]<br>• Adaptive active damper [52]<br>• Adaptive controller gain tuning [53]. |

A voltage feedforward element, based on the voltage measured at the PCC, is usually added to the control scheme of inverters to achieve faster dynamic performance. However, under weak grid conditions, the voltage feedforward term could make the inverter more susceptible to instability [24], [43], [44]. Thus, the voltage feedforward term needs to be modified to ensure stability in weak grids, and some recent works in [1], [24], [45] feedforward term was modified to improve the stability in weak grids. In [46], the feedforward term was modified through a delay compensation function, where the delay compensation transfer function was derived for one and a half sampling period delay. The inverter displayed improved robustness to grid impedance variation with the modified feedforward term. A capacitor voltage feedforward term was also considered in [47], with the feedforward term added to the control scheme through a first-order filter and a constant gain term. The gain term can be designed to make the inverter robust under weak grid conditions. In [48], PCC voltage feedforward was employed through a filter and a gain block. It was shown that for the modified voltage feedforward with a suitable value of the gain block between 0 and 1, the performance of the inverter displayed improvements under weak grids. Increasing the grid-side filter inductance, $L_2$ of LCL filter could result in stable inverter operation at weak grids as demonstrated analytically in [49]. The authors developed a virtual inductance feedforward control strategy emulating the characteristics of extra inductance in the grid-side filter inductance. The value of virtual inductance must be selected carefully to ensure the stable operation of the inverter in weak grids [49]. Although each of the feedforward techniques described above demonstrated improved performance in weak grids, they all have one common shortcoming. Each one of these techniques requires manual tuning of the feedforward term. As a result, they will be challenging to implement in practical applications where the grid impedance will be unknown.

The adaptive techniques use the same principles as the previously described techniques to improve inverter stability in weak grids, though they have an adaptive element, which makes them suitable for inverters [50], [51]. An adaptive control scheme that relies on the real-time estimation of grid impedance has been developed in [51] to enhance the stability of inverters. In [52], an active damper was added to the system, which essentially introduces an additional resistive term in the inverter circuit that can be varied adaptively to make the inverter more robust against changes in grid impedance. The active damper was realized using an additional low-power single-phase inverter, which ensures stability without adding any complexity to the controller of the actual inverter, however, it requires extra circuitry contributing to additional size and cost. In [53], the gain of the voltage feedforward path was updated adaptively to enhance stability in weak grids. However, the technique presented in [53] required grid impedance estimation, implemented using a band-pass filter, the parameters of which were changed manually.

## III. Ancillary Services Via Grid-Interactive Inverter

Grid-interactive inverters encounter various abnormal grid conditions. Asymmetrical and symmetrical voltage sags are the most common anomalies [1], [26]-[27]. As the number of DER-based inverters is increasing in the power grid, they are required by the utilities to stay connected to the system under such voltage sags, known as low voltage ride through (LVRT). Smart inverters must be capable of identifying such grid anomalies, stay connected to the grid, and provide ancillary services [28]-[29]. The ancillary services of the inverter are reviewed separately for symmetrical and asymmetrical faults below:

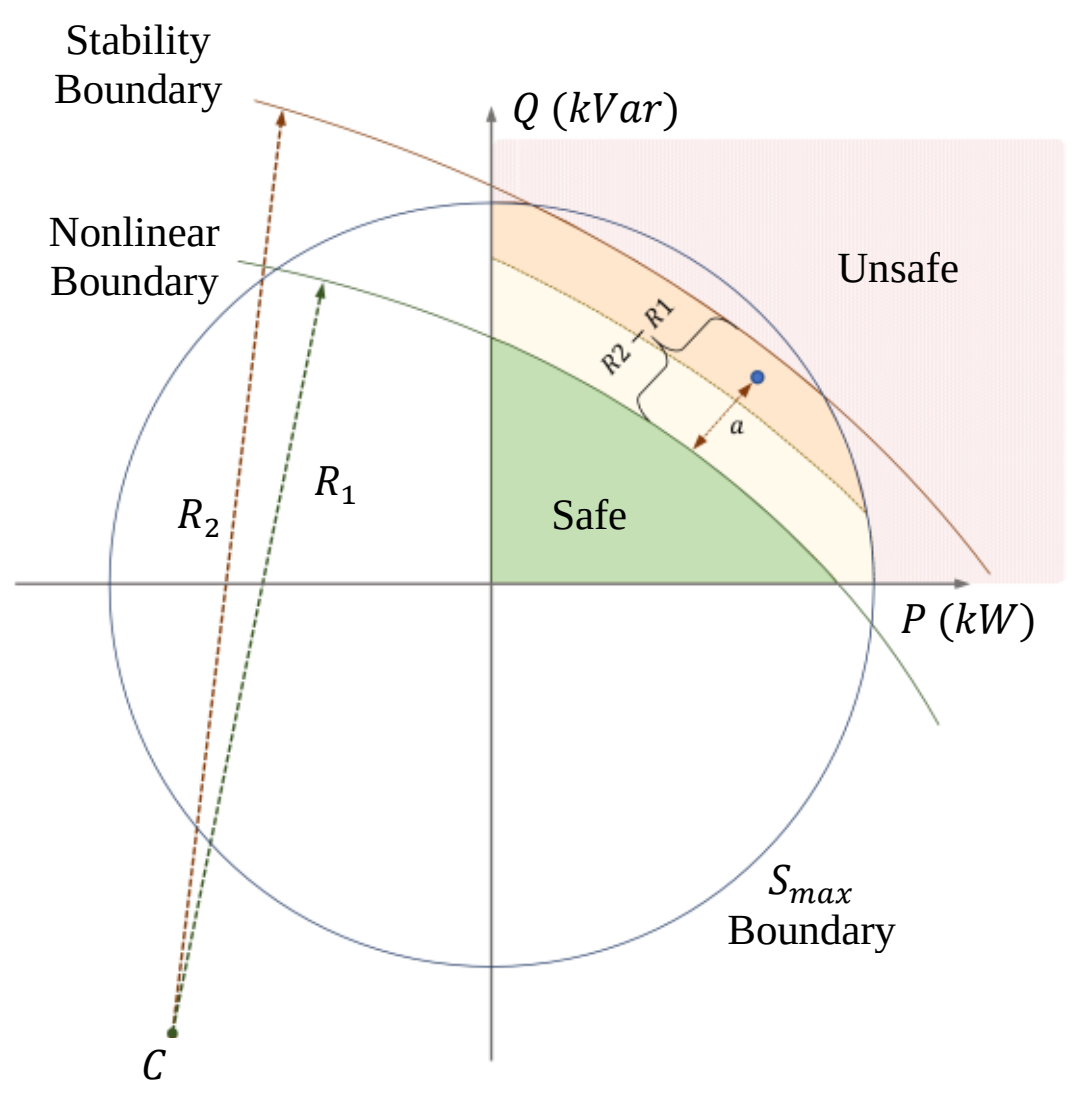

Fig. 3: Linear and stable operating regions of a grid-interactive inverter.

### A. *Ancillary Services During Symmetrical Anomalies*

Symmetrical voltage sag happens when the utility has symmetrical faults. According to previous utility standards, inverters were required to disconnect from the grid during voltage sags [54]-[55]. However, as the utilization of inverter-based renewable sources increases, a sudden loss of power from all the inverter-based DGs could result in a significant consequence than the voltage sag itself. Therefore, inverters are required to stay connected to the grid under voltage sags. Furthermore, according to some European grid codes [56] and IEEE std. 1547-2018 [57], the inverters are required to provide reactive power support during the grid voltage sags. If the voltage sag exists too long, the inverter should be disconnected from the grid. One important thing to note is that the active current fed to the grid should be adjusted as the reactive current is increased to avoid violating the maximum current limit of the inverter shown as a circle with a radius of $S_{max}$ in Fig. 3, which would otherwise trip the inverter [58], [59]. Besides this, the amount of active and reactive power delivered from the inverter's terminal to the grid can be formulated as [60],

$$S = \frac{|V_{inv}||V_{th}|e^{-j(\delta-\theta_Z)} - |V_{th}|^2 e^{j(\theta_Z)}}{|Z|} \quad (1)$$

where, $V_{inv} = |V_{inv}|\angle\delta$ and $V_{th} = |V_{th}|\angle 0$ are respectively the inverter output and grid Thevenin voltages, $\delta$ is the angle difference between the inverter and grid voltages. $Z = (Z_f + Z_{th})\angle\theta_Z$ is the impedance seen from the inverter's output terminals. Herein, the inverter's fundamental output voltage can be represented as $v_{inv} = (k_{m,i})V_{dc}\sin(\omega t + \delta)$, where $k_{m,i}$ is the DC-bus utilization factor. For three-phase grid-interactive inverters using sinusoidal PWM (SPWM) and space-vector PWM or SPWM+3rd harmonics techniques, $k_{m,1}$ becomes,

$$\begin{cases} k_{m,1} = \dfrac{\sqrt{3}}{(2\sqrt{2})}\, m & (SPWM) \\ k_{m,1} = \dfrac{(1.1547)\sqrt{3}}{(2\sqrt{2})}\, m & (SVPWM) \end{cases} \quad (2)$$

where, $0 < m \leq 1$ denotes the modulation index. When the inverter operates in a nonlinear region, this utilization ratio can be violated due to over-modulation, also known as pulse dropping, and more harmonics can be observed since the system operates in a nonlinear region [1], [11]. The maximum limit (boundary of instability) in which an inverter is capable of utilizing the DC-bus voltage can be defined using $k_{m2} = m(4\sqrt{3})/(2\pi\sqrt{2})$ because the RMS value of the fundamental component of a square waveform is $4/\pi$. Based on the Fourier series, a voltage source inverter (VSI) cannot utilize the DC-bus voltage more than $k_{m2}$ ratio, in theory [36]. Eq.1 can be represented as a circle with the radii for linear and nonlinear operating boundaries, $R_i$, and center, $C$, can be written as

$$(P - Real(C))^2 + (Q - Imag(C))^2 = (R_i)^2 \quad (3)$$

where,

$$\begin{cases} C = \dfrac{-Z|V_{th}|^2}{|Z|^2} = -\dfrac{|V_{th}|^2 \cos(\theta_Z)}{|Z|} - j\dfrac{|V_{th}|^2 \sin(\theta_Z)}{|Z|} \\ R_i = \dfrac{k_{m,i} V_{dc} |V_{th}|}{|Z|} \end{cases} \quad (4)$$

where, $R_i, i \in \{1,2\}$, and $R_1$ denotes the radius of the linear (safe) operation boundary, and $R_2$ represents the radius of instability boundary, see Fig. 2. In Fig. 2, the safe area is represented as the overlap area of the inner circle with $S_{max}$ boundary. The area outside the inner circle with a radius $R_1$ enclosed by the intersected area of the $S_{max}$ circle with the outer circle with radius $R_2$ defines the nonlinear operating region. Grid-supporting inverters extract the maximum available power from the intermittent source, e.g., PV arrays, and feed to the grid. In DG units, voltage source inverters (VSIs) are typically utilized, but current source inverters (CSIs) can also be used [31]-[33]. A CSI, unlike a VSI, is a boost inverter that may be fed by a parallel connection of small dc sources, improving the reliability and availability of DG units [61].

### B. *Ancillary Services During Asymmetrical Anomalies*

Due to asymmetrical faults in the utility and uneven distribution of single-phase loads in the distribution network, asymmetrical voltage sag could develop [36], [62]. The asymmetrical faults can be categorized as single-phase to ground fault, two-phase to ground fault, and phase to phase fault [25]. In such situations, an inverter can assist the grid by providing negative-sequence compensation services in addition to positive-sequence reactive power assistance [63]. The voltage imbalance factor (VUF) of the PCC voltages can be used to identify asymmetrical voltage sag. There would be no need for additional sensors because PCC voltages are already monitored. Two current control loops can be added to the control scheme for providing negative-sequence compensation to the grid currents, irrespective of whether the control scheme is based on stationary reference frame components or synchronous reference frame components [1], [64]. The newly added control loops will operate based on the negative-sequence components of the grid current as shown in Fig. 3, and the setpoints of the controllers can be set at any value other than zero to mitigate some of the oscillations. A common-mode injection scheme can be added to the control scheme to limit the inverter references in the linear modulation region while providing ancillary services [42]. Therefore, a smart inverter equipped with a common-mode injection scheme can maximize its dc-bus voltage utilization.

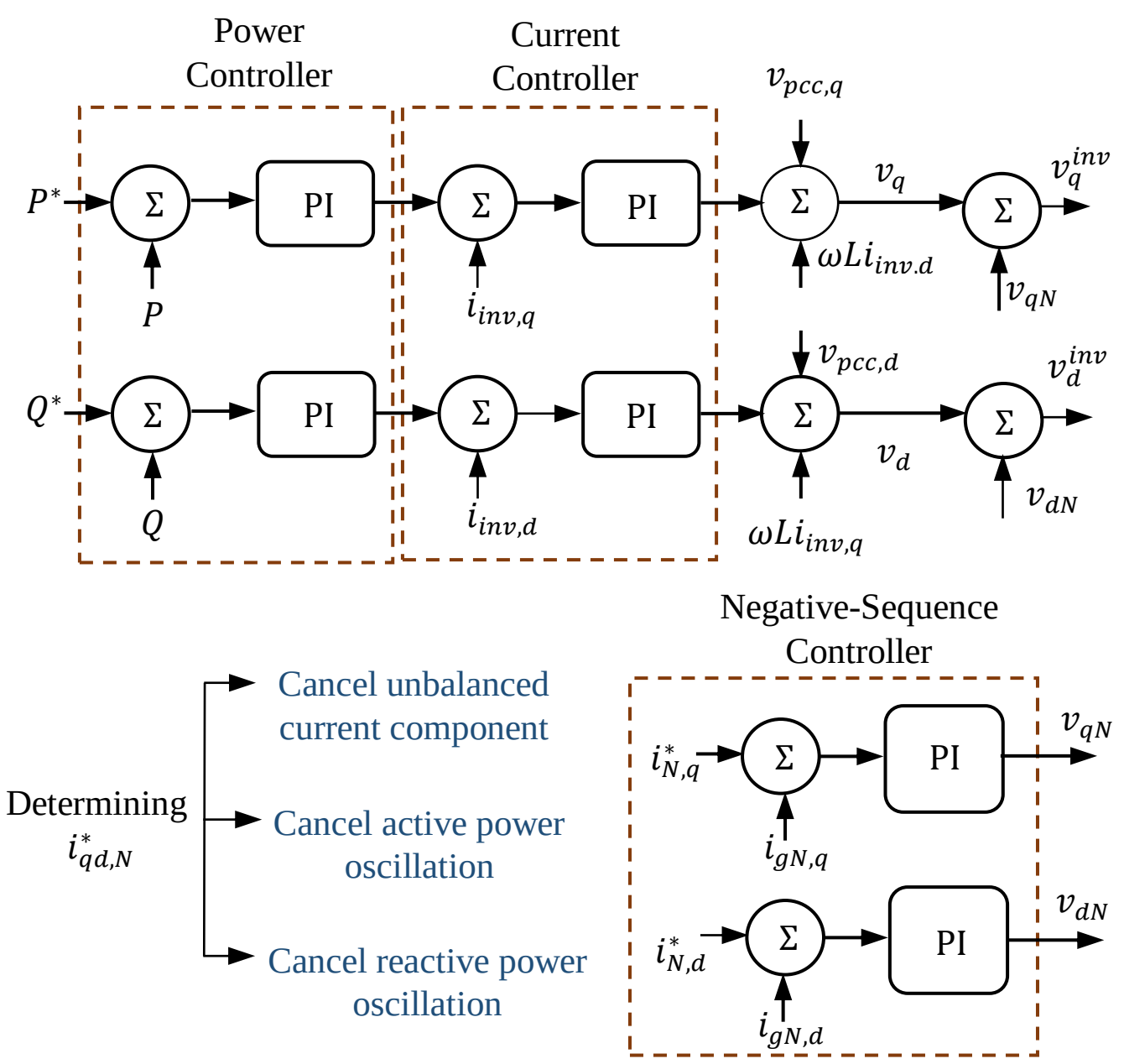

Fig. 4. Control block diagram and decision chart for negative-sequence controller of inverter during grid-supporting mode of operation [1].

Phase-angle detection in abnormal three-phase systems is still a challenging task for enhancing the stability margin of grid-tied smart inverters. To increase the inverters resiliency, advanced phase-angle detection schemes are needed. A direct phase-angle detection method has been presented in [65]. This feature consists of a direct instantaneous phase-angle detector capable of accurately detecting the phase-angle of the desired reference voltage under unbalanced grid conditions. Accurate phase-angle detection is a vital function of the self-learning inverter as it is used in the synchronization and controller scheme computations of the inverter [66].

## IV. Security of Grid-Interactive Inverter

Recent developments in communication and internet technologies allow advanced features in grid-interactive inverters. The inverters connected to communication or cyber network could be in danger of being hacked. This section discusses grid-interactive inverters as a part of a cyber-physical system and a review of possible cyber-attacks and security assessments.

### A. *Cyber-Physical Devices and Possible Attack scenario*

Grid-interactive inverters in a cyber-physical power grid and possible cyberattack scenarios targeting the inverter are shown in Fig. 5. Inverter control systems need system parameters to perform control actions and achieve desired operating conditions. These parameters can be received as external data. For example, the grid-interactive inverter controllers can obtain measurement data from smart devices, power setpoints from utility services, forecasting data from weather services, etc. Communication with multiple parties through the cyber network allows autonomous interactions and information exchange; however, it entails more surface for harmful activities, specifically when the communication protocols are insecure, operating system software, and user credential information are outdated [4], [67],[68]. These activities can be represented as intentional and unintentional unsafe events that can impede the normal operation of the inverters. For instance, an authorized utility operator can unintentionally send unsafe $PQ$ setpoints that can operate the inverter beyond its capability limit or cause pulse dropping and high total harmonic distortions (THD) [69]. On the contrary, an unauthorized user can intentionally change measurement data received from an external sensor. The most challenging aspect of operating a communication-enabled device is to ensure security since hackers want to jeopardize the operation of the inverters. In particular, as inverter-based has low inertia, they are vulnerable to sudden changes in operating point that can cause voltage sags or swells and the instability of the controllers that can lead to cascaded inverter trips [70], [71]. Insecure communication protocols, outdated software, weak user credentials allow hackers to breach the inverter system via the communication link and perform the data modifications. Possible attack scenarios are security certificate proof, brute force credentials (BFC), man-in-the-middle (MITM), denial of service (DoS), replaying, eavesdropping, etc., as discussed in [10].

### *B. Detection and Protection Against Probable Cyber Attacks*

In recent years, cyber-attack identification and prevention have been a trending research area. Despite various researches being ongoing, no investigation can guarantee the secure operation of devices that are connected to the cyber-physical system. In literature, numerous cyberattack detection algorithms have been developed, which include advanced communication protocol, intensive computation process, i.e., machine-learning algorithm to ensure security against cyber-attacks.

Some research includes knowledge-based techniques to ensure the safe operation of the inverters [71]. Hackers can bypass security protocols, authentication restrictions, user interface firewalls, and other barriers by using their coding skills and performing cyber-attacks. The literature in [9], [72] emphasizes data transfer security to provide security against cyberattacks. These solutions, however, are inadequate to address the security issue with smart inverters. Hence, protection beyond network security is required to ensure the inverter's safe operation. The safe operation can be accomplished by creating techniques based on system and device-level data. System-level techniques use system data sent from the neighboring cyber-physical devices and controllers to investigate whether there is a cyber-attack. In [73]-[74], a data-driven method such as machine learning is used to ensure system-level security.

### *C. Inverter Self-Security*

Despite the fact that many investigations on software and system-level security have been reported, device-level detection is suggested in [10]. A model-based self-security algorithm was first developed in [71]. The authors underlined that model-based approaches could be included in the inverter control circuit to inspect the validity of incoming power setpoints. The model-based device-level security technique is depicted in Fig. 6. In [69], [71], three reference model has been utilized to guarantee the inverter security. The utilized reference models include maximum inverter capability, $S_{max}$ As shown in Fig. 2, instability boundary, utilizing the concept from [42] and [58], and examine dynamic performance using the reduced-order digital-twin model to represent the actual dynamic behavior and to achieve faster operations [69], [71]. In addition to the models discussed in [71], IEEE 1547-2018 standards requirements such as high-voltage ride-through (HVRT) and low-voltage ride-through (LVRT) can be applied as an additional knowledge-based model to ensure safe operation. Therefore, a self-security algorithm can ensure a secure operation by detecting anomalies in the incoming data. When adequate information is provided to the inverter, it can learn its dynamic performance, stability boundaries, and maximum capabilities. The inverter can safeguard itself and the utility grid from abnormal operation when the reference model is appropriately formed.

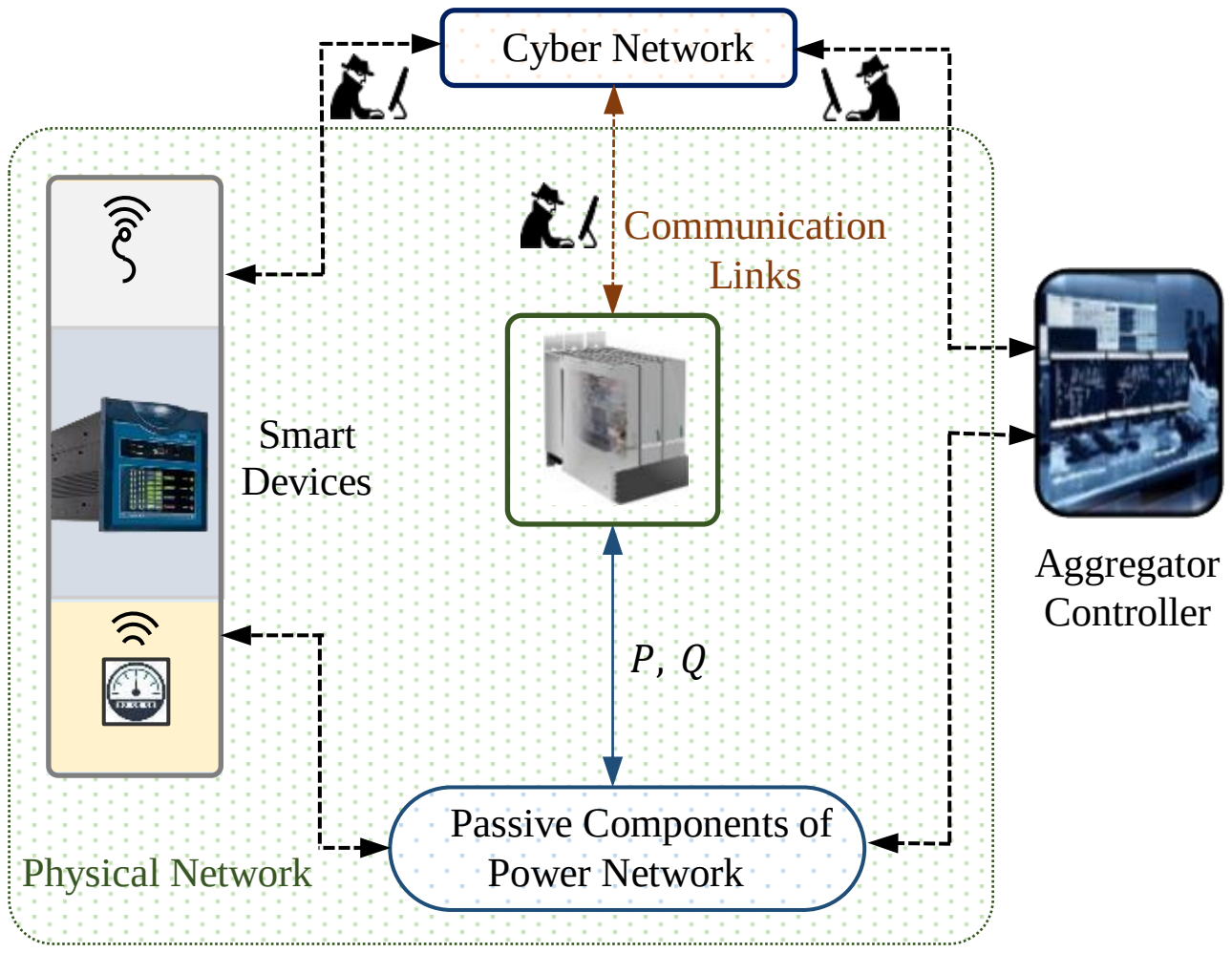

Fig. 5. Power network connected smart inverters and possible cyberattack scenarios.

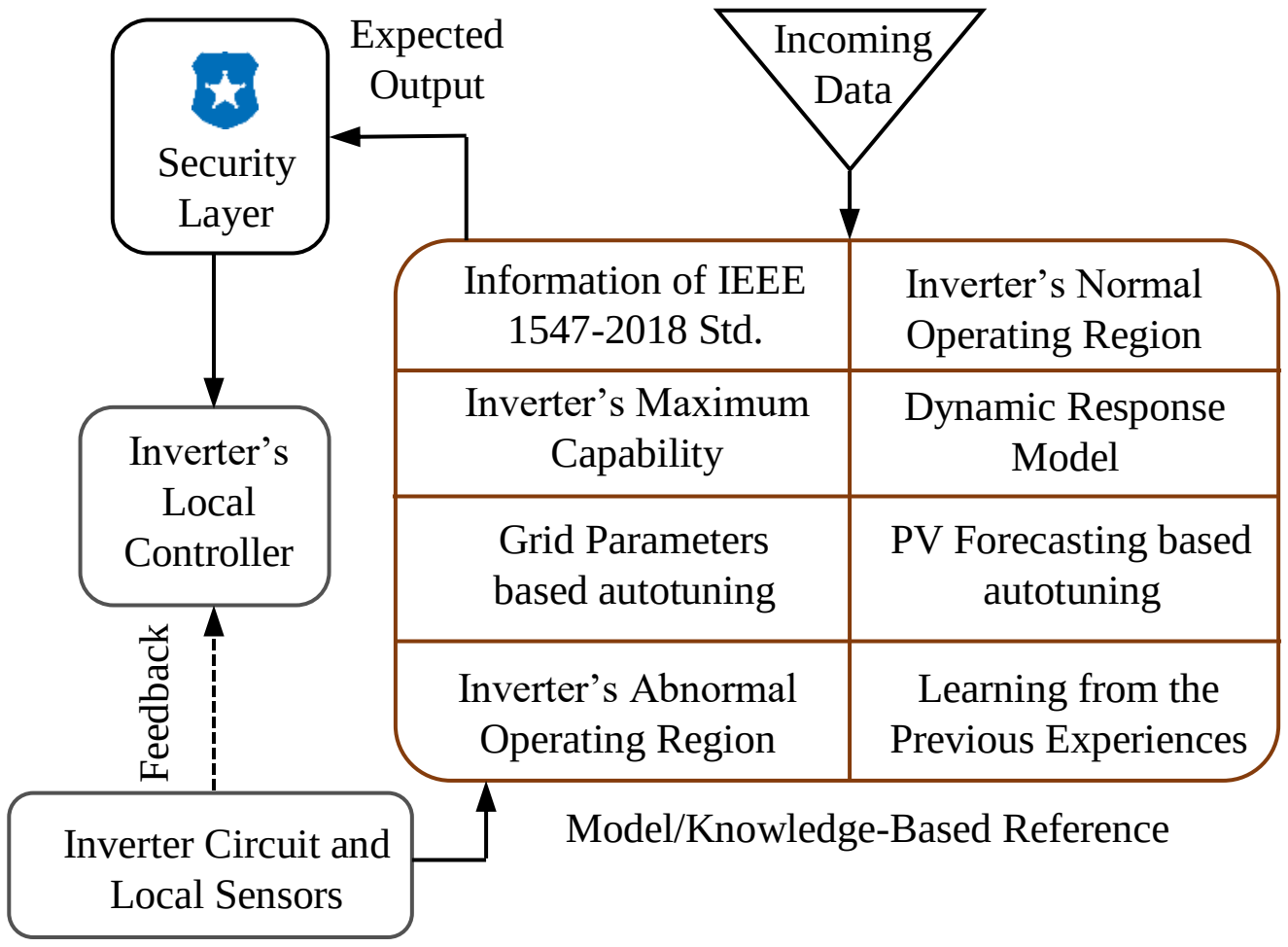

Fig. 6. Reference model-based device level security techniques.

## V. Multi-Inverter Operation

The coordinated operation between the multiple inverters ensures a more secured environment under adverse grid conditions and external attacks. Seamless synchronization is necessary for multiple inverters operating in a system. Most synchronization techniques address the connection of an inverter or a microgrid to a power grid. For example, synchronization methods for seamless transitions between grid-tied and stand-alone modes of a single inverter are reported in [75]-[79]. The microgrid (re)connection to a power grid is reported in [80], where the inverters follow the grid frequency using secondary control schemes such that the frequency of the grid and microgrid remains unchanged during any synchronization process. In most reported studies, the inverters change their mode of operation to grid-following after the process of synchronization [81]-[85]. In [86], communication infrastructure is implemented for synchronization, and droop controls are not applied. The synchronization issue of an inverter as part of a microgrid fed by multiple inverters is addressed in [87]. A seamless synchronization method of inverters for microgrid operation has been developed. The strategy ensures seamless and secure connection/reconnection for new inverters added to the system, as well as ensuring equal power-sharing of incoming sources [88].

PLL malfunctioning can cause an inverter to lose the synchronization, consequently shutting down the inverter operation, leading to significant power loss. In literature, numerous synchronization methods have been developed for abnormal grid circumstances [89], including advanced PLL designs [90], compensating impedance, and feedforward design to avoid such situations [91]-[93]. Several PLL-free fast and robust phase-angle detection techniques were developed in [94]-[95]. These PLL-less phase-angle detection methods are compatible when traditional generation units are involved. To resolve the problem of PLL misfunctioning, a cooperative strategy has been developed in [66].

Single-phase microgrids feeders can be disconnected and operate as stand-alone microgrids with sufficient DGs after losing the utility. Placement of interphase AC-AC inverters between islanded single-phase feeders have been recommended for power flow control between the feeders [65],[97]. Here, the three single-phase feeder's phase sequences remain the same as actual $abc$-sequence [65], [97]. As a part of the coordinated operation, a strategy for three single-phase to one single phase microgrid development to fulfill the load demands has been developed in [98]-[100].

## VI. Conclusion

This paper highlighted some current contributions on enhancing the stability and self-security of grid-interactive inverters. In particular, stability of grid-interactive inverters in weak grids has been discussed for different LCL filter parameters. The recommendation is to add a feedforward path in the conventional control scheme that emulates an increased grid-side filter inductance effect. Grid-interactive inverters can also utilize the maximum dc-bus voltage by adding a nonlinear common-mode reference single into the PWM generator while providing harmonic and negative sequence compensations. Basically, the linear modulation region of inverters can be expanded by adding the common-mode reference signal. Autonomous security from cyberattacks is another recent research area that has been highlighted in this paper. The requirement for device-level security measures has been addressed. The use of reference-model techniques has been recommended to ensure the safe operation of the inverter during anomalies and cyber-attacks. Also, this paper has briefly discussed a multiple-inverter collaborative operation for smooth power-sharing and synchronization.